\documentclass[prb]{revtex4}

\usepackage{graphicx}
\usepackage{dcolumn}
\usepackage{bm}
\usepackage{pst-all}
\usepackage{subfigure}
\usepackage{multirow}

\def\e{\begin{equation}}
\def\f{\end{equation}}
\def\*{^{\displaystyle*}}

\def\x.{\displaystyle{{}^\times}\llap{${}_\cdot$}}
\def\.x{\,\displaystyle{{}^\cdot}\,\llap{${}_\times$}}
\def\%#1{\mbox{\boldmath $#1$}}
\def\=#1{\overline{\overline #1}}

\def\E{\epsilon}
\def\M{\mu}

\def\.{\cdot}
\def\x{\times}
\def\##1{{\mathbf#1\mit}}
\def\_#1{{\mathbf#1\mit}}

\def\Re{{\rm Re\mit}}

\def\l#1{\label{eq:#1}}
\def\r#1{(\ref{eq:#1})}
\def\am{\left(\begin{array}{c}}
\def\amm{\left(\begin{array}{cc}}
\def\a{\end{array}\right)}

\def\vecs{\mathsf}
\def\vec{\mathbf}

\def\x{\times}

\def\=#1{\overline{\overline #1}}

\def\M{\mu}

\def\E{\epsilon}

\def\A{\alpha}

\begin{document}

\title{Backward-wave regime and negative refraction in chiral composites}


\author{S. Tretyakov, A. Sihvola, and L. Jylh\"a}
 \email{Sergei.Tretyakov@tkk.fi}
\altaffiliation{Department Electrical and Communications Engineering\\
Helsinki University of Technology\\
P.O. 3000, FI-02015 TKK, Finland}

\begin{abstract}

Possibilities to realize a negative refraction in chiral composites in in
dual-phase mixtures of chiral
and dipole particles is studied. It is shown that because
of strong resonant interaction between chiral particles (helixes) and dipoles, there is a stop
band in the frequency area where the backward-wave regime is
expected. The
negative refraction can occur near the resonant
frequency of chiral particles. Resonant chiral composites may offer a root to
realization of negative-refraction effect and superlenses in the optical region.

\end{abstract}

\maketitle


\bigskip

\bigskip

\section{Introduction}

It is well known that negative refraction takes place at an interface
between a usual isotropic medium (vacuum, for example) and a
material with negative permittivity and permeability (called Veselago medium,
double-negative material, or backward-wave medium).
Recently, a lot of efforts have been devoted to realization of
backward-wave materials, because the negative refraction effect offers a possibility to
create super-resolution imaging devices (among other potential applications).
The known realizations are based on the use of metal inclusions of
various shapes, especially split rings, needed to realize negative permeability.
Creation of strong artificial magnetic response, especially in the optical region, is
a big challenge, which we have to face if we want to realize negative
refraction and superlens for optical applications. In view of
this problem, various alternative approaches to create backward-wave media
have been considered in the literature. In particular, such effects can exists in
more complex materials --- chiral media --- and, what is the key advantage,
backward-wave regime can be in principle realized even if the medium has very weak or
no magnetic properties. Thus, it appears that using chiral media one could
realize negative refraction in the optical region without the need to
create artificial magnetic materials operational in that frequency range.

The physics of the effect of backward waves in chiral media is
very simple: The propagation constants of two eigenwaves in
isotropic chiral media equal $\beta=(n \pm \kappa)k_0$, where
$n=\sqrt{\epsilon\mu}$ is the usual refractive index, $\kappa$ is
the chirality parameter, and $k_0$ is the free-space wavenumber
(see, e.g., \cite{chibi}). Near the resonance of electric or /and
magnetic susceptibilities the refractive index $n$ can become
smaller than the chirality parameter $\kappa$. It means that one
of the two eigenwaves is a backward wave, because its phase
velocity is negative but the energy transport velocity is
positive. At an interface between a usual isotropic material and
such medium negative refraction takes place for this polarization
(waves of the other polarization refract positively). The earliest
publication where a possibility for such effects was established
was probably paper \cite{bokut}. In that paper, a spiral model for
a chiral optical molecule and the Lorentz dispersion model for the
permittivity was used, and a formula for the frequency range of
negative refraction was derived. Single-phase chiral substances
were considered, and magnetic properties of the medium were
neglected. Much more recently, backward waves and negative
refraction were studied in \cite{sergei1}, with the emphasis on
the limiting case when both $\epsilon$ and $\mu$ tend to zero
(this medium was called {\itshape chiral nihility}. The simplified
antenna model for chiral inclusions \cite{antenna} was used to
estimate the material parameters of mixtures of metal helices with
the desired parameters. Possibility for backward-wave regime in
chiral materials was also indicated in conference presentation
\cite{Dakhcha1}. In paper \cite{pendry1}, instead of chiral
nihility, a two-phase mixture was introduced. One phase is a
non-dispersive chiral material, and another phase consists of
resonant dipole particles. It has been assumed that when the
dipoles resonate, $\sqrt{\epsilon}$ becomes smaller than $\kappa$,
and the material can support backward waves. These results show
that the use of chirality is a very exciting new opportunity to
realize negative refraction and related effects in the optical
region in effectively uniform media (the characteristic dimensions
in the material can be much smaller than the wavelength).

Chiral media have been very intensively studied in the past years, see e.g.
\cite{chibi,serdyukov1,lakhtakia2,priou2,mariotte1}.
However, it is interesting, that although the possibility for this effect was
published in the former Soviet Union \cite{bokut}, it was not known
in the West until very recently. The authors of monograph \cite{chibi} thought that both
eigenwaves in chiral media should be forward waves, and formulated a corresponding
restriction for the material parameters [See Eq.~(2.176) on page
51]. Recent studies have shown, that one of the eigenwaves in a
chiral nihility is indeed a backward wave \cite{sergei1}.

In this paper we study eigenwaves propagating in single- and dual-phase
chiral mixtures accurately taking into
account resonant properties of chiral particles and of resonating
electric dipoles particles and electromagnetic interaction between phases.
We identify the effects that can lead to realization of backward-wave
regime and negative refraction using chiral composites.

\section{Backward waves in chiral media and enhancement of evanescent fields}

 The bi-isotropic constitutive relations read
\begin{equation}\label{bi}
\am {\vec D} \\ {\vec B} \a = \amm \E & \xi \\ \zeta & \M \a \am
{\vec E} \\ {\vec H} \a = {\vecs M} \cdot \am {\vec E} \\ {\vec H}
\a
\end{equation}
where the material matrix $\vecs M$ contains the four scalars
$\E,\xi,\zeta,\M$.
Here we will consider only isotropic reciprocal media, in which
the following condition holds \cite{chibi}:
\e \zeta=-\xi=j\kappa\sqrt{\epsilon_0\mu_0}\l{rec}\f
Such media are called chiral, and $\kappa$ is the chirality parameter. In lossless
media $\kappa$ is a real number.

Considering electromagnetic field in
homogeneous chiral regions it is very convenient to introduce new
field variables $\_E_\pm$ and $\_H_\pm$ that are the following
linear combinations of the fields:
\e \_E_+={1\over 2}(\_E-j\eta\_H),\qquad\_E_-={1\over
2}(\_E+j\eta\_H)\f
\e \_H_+={1\over 2}\left(\_H+{j\over \eta}\_E\right),\qquad\_H_-=
\left(\_H-{j\over \eta}\_E\right)\f
where $\eta=\sqrt{\mu/\epsilon}$. Vectors $\_E_\pm$ are called wavefield vectors.
The advantage of introducing the new variables comes from the fact that these
new vectors satisfy the Maxwell equations in equivalent isotropic
non-chiral media. This allows to use the known solutions for fields in
simple isotropic medium to construct solutions for fields in
chiral media.

In uniform regions the wavefield components "see" equivalent simple
isotropic media with the equivalent parameters
\e \epsilon_\pm=\epsilon(1\pm \kappa_r),\qquad \mu_\pm=\mu(1\pm \kappa_r)\f
where $\kappa_r=\kappa/\sqrt{\epsilon\mu})$ \cite{chibi}.
The wavenumbers of the two eigenwaves
read $ k_\pm=k(1\pm \kappa_r)$,
or the refractive indices of the two equivalent media are $n_\pm=\sqrt{\epsilon\mu}
\pm \kappa$.

\begin{figure}[h]
\includegraphics[width=0.47\textwidth]{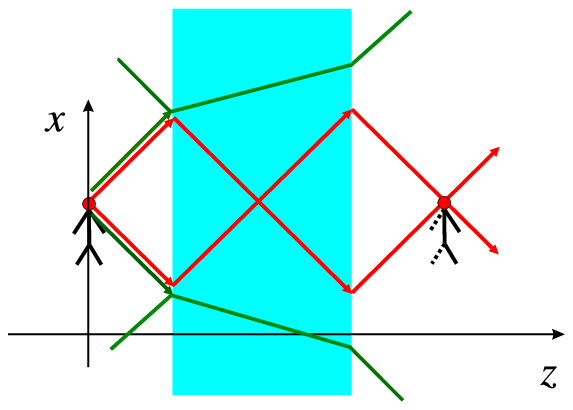}
\includegraphics[width=0.52\textwidth]{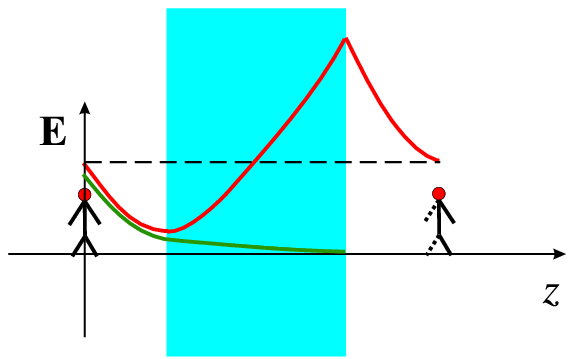}
\caption{An illustration of negative refraction and subwavelength
focusing by a chiral slab.}
\label{illustration}
\end{figure}

Suppose that in some frequency region
\e \Re\{\kappa\} > \Re\left\{\sqrt{\epsilon\mu}\right\} \l{condition}\f
 In this case
 one of the eigenmodes is a backward wave.
Actually the two eigenmodes $\_E_+,\_H_+$ and $\_E_-,\_H_-$ are
plane right- and left-circularly polarized waves. We see that for
one of these two polarizations a slab of chiral material [when
\r{condition} is satisfied] behaves as a slab of an isotropic
medium with negative effective parameters (Veselago medium). The
known phenomena of negative refraction and subwavelength focusing
will take place for waves of this polarization, see an
illustration in Figure~\ref{illustration} (These pictures
have been drawn by S. Maslovski.). Numerical simulations of
focusing effect have been published in \cite{sailing_focusing}.


The condition for creation of a perfect image for one of the two
circular polarizations read
\e \epsilon_-=\epsilon(1- \kappa_r)=-\epsilon_0,\qquad \mu_-=\mu(1- \kappa_r)=-\mu_0 \qquad
(n_-=-1)\f

\section{Mixture of chiral and dipole particles}

In \cite{pendry1} it was assumed that once knowing the effective
permittivity of a mixture of dipoles and the effective $\kappa$ of
a chiral mixture, the propagation constants of wave in a composite
material which contains both types of inclusions
could be calculated by a simple substitution to
\begin{equation}
k_\pm=(\sqrt{\epsilon_{\rm eff}} \sqrt{\mu_{\rm eff}}\pm
\kappa_{\rm eff})k_0 \label{beta}
\end{equation}
where $k_0$ is the wave number in free space.
 However, this approach does not take into
account electromagnetic coupling between dipoles and helixes,
which is rather strong near resonant frequencies of at least one
phase. In this section, we introduce effective material parameters
which are calculated taking this coupling into account. We assume
that the mixture is built up by randomly distributed dipoles and
helices in a uniform matrix with the permittivity of air. Another
possibility is to construct a regular lattice of symmetrically
positioned inclusions, so that the overall response is
isotropic. In both cases the scattering losses from
individual inclusions are supposed to be compensated, and the material
can have rather low loss factor, determined only by dissipation in
particles. The mixture is assumed to be dilute, in other words particles
are not in each other near field, and the Clausius-Mossotti model
can be used.

In the following formulation, let us use normalized field
quantities: in terms of the ``ordinary'' electric and magnetic
fields and flux densities $\cal E,H,D,B$ with units V/m, A/m,
As/m$^2$, Vs/m$^2$, respectively, we deal with fields and fluxes
that are renormalized in order to have homogeneous units in each
of them:
\begin{equation}
{\vec E}=\sqrt{\E_0}\,{\cal E}, \quad {\vec H}=\sqrt{\M_0}\,{\cal
H}, \quad {\vec D}=\frac{{\cal D}}{\sqrt{\E_0}}, \quad {\vec
B}=\frac{{\cal B}}{\sqrt{\M_0}}
\end{equation}
with the free-space parameters $\E_0,\M_0$. This leaves the
materials parameters in Equation~(\ref{bi}) dimensionless. Also,
all four renormalized field quantities carry the dimension of
square root of energy density: $\sqrt{{\rm VAs/m}^3} = \sqrt{\rm
J/m^3}$.

\subsection{Effective parameters}
\label{mixing}

Consider a mixture where randomly oriented helices and randomly
oriented dipole resonators float in neutral matrix (air). Let the
polarizability matrix contribution of one element in the random
distribution of helices be
\begin{equation}
{\mathsf A} = \amm \A_{ee} & \A_{em} \\ \A_{me} & \A_{mm} \a
\end{equation}
where the averaging over spatial orientation has been included, in
other words,
\begin{equation}
\A_{ee} = \frac{1}{3} (\A_{ee,\rm x} +\A_{ee,\rm y} +\A_{ee,\rm
z}), \quad \A_{em} = \ldots
\end{equation}
Note also that we can assume reciprocity, in other words
\begin{equation}
\A_{me} = -\A_{em}
\end{equation}
[see Eq.~\r{rec}].
Of course, for lossless helices, $\A_{me}$ and $\A_{em}$ are pure
imaginary.

Let the averaged polarizability matrix of the dipole resonators be
\begin{equation}
{\mathsf B} = \amm \beta_{ee} & 0 \\ 0 & 0 \a
\end{equation}
with the similar averaging included. Here we have assumed that in
addition to the vanishing magnetoelectric contribution, the
magnetic response can be neglected. (This assumption is not
necessary but makes the point easier to demonstrate that we can
have different band behaviors for the macroscopic permittivity and
chirality responses.)

Using the
methods described in
\cite{ari1,serdyukov1}, the Maxwell Garnett formula in a matrix form for the effective
parameters of the mixture can be written in the following way:
\begin{equation}
({\mathsf M}_{\rm eff} - {\mathsf I})\cdot ({\mathsf M}_{\rm eff}
+ 2 {\mathsf I})^{-1} = \frac{1}{3} (n_1{\mathsf A} + n_2 {\mathsf
B})
\end{equation}
where $n_1$ and $n_2$ are the number densities of the helices and
dipoles, respectively. The explicit formula for the effective
parameter matrix
\begin{equation}
{\mathsf M}_{\rm eff} = \amm \E_{\rm eff} &  \xi_{\rm eff} \\
 \zeta_{\rm eff} & \M_{\rm eff} \a
\end{equation}
reads
\begin{equation}
{\mathsf M}_{\rm eff} = {\mathsf I } + \left( {\mathsf I} -
\frac{1}{3}(n_1{\mathsf A} + n_2 {\mathsf B})\right)^{-1} \cdot
 (n_1{\mathsf A} + n_2 {\mathsf B})
\end{equation}

The resulting parameters read
\begin{eqnarray}
\E_{\rm eff} & = & 1 + \frac{3n_1\A_{ee} + 3n_2\beta_{ee}
- n_1^2 (\A_{ee}\A_{mm}+\A_{em}^2) - n_1\A_{mm} n_2\beta_{ee}}{3\cdot {\rm DEN}} \label{eps} \\
\xi_{\rm eff} & = & \frac{n_1\A_{em}}{{\rm DEN}} \label{xi}\\
\zeta_{\rm eff} & = & -\frac{n_1\A_{em}}{{\rm DEN}} \label{zeta}\\
\M_{\rm eff} & = &  1 + \frac{3n_1\A_{mm} - n_1^2
(\A_{ee}\A_{mm}+\A_{em}^2) - n_1\A_{mm} n_2\beta_{ee}}{3 \cdot{\rm
DEN}} \label{mu}
\end{eqnarray}
where the common denominator is
\begin{equation}
{\rm DEN} = 1 - \frac{n_1\A_{ee}}{3} - \frac{n_1\A_{mm}}{3} +
\frac{n_1^2(\A_{ee}\A_{mm}+\A_{em}^2)}{9} -
\frac{n_2\beta_{ee}}{3} + \frac{n_1\A_{mm} n_2\beta_{ee}}{9}
\end{equation}
Obviously the material parameters depends on  polarizabilities in
a very complicate way, such that every material parameter depends
on all the polarizabilies of both types of inclusions.

\subsection{Polarizabilities of chiral inclusions}

Analytical models of polarizabilities of chiral inclusions are
well known in the literature, see e.g. \cite{antenna,serdyukov1,priou2,mariotte1}.
All the polarizabilies (electric, magnetic, and magneto-electric) have
the resonant behaviour with the same resonant frequency
which corresponds to the resonance of the whole particle.
At low frequencies (well below the resonance), the electric polarizability tends to
a constant, chirality parameter tends to zero linearly with the frequency, and
the magnetic polarizability is proportional to frequency squared:
\e \alpha_{em}\sim j(ka) \alpha_{ee},\qquad \alpha_{mm}\sim -j(ka) \alpha_{em}\sim  (ka)^2
\alpha_{ee}\l{hie}\f
where $a$ is the characteristic particle size.
This brings us to the following generic model of the
frequency dispersion of the polarizabilities:
\begin{equation}
 \alpha_{em}={jAx\over{1-x^2+j\alpha x}},\qquad
\alpha_{ee}={B\over{1-x^2+j\alpha x}},\qquad
\alpha_{mm}={C x^2\over{1-x^2+j\alpha x}}
\end{equation}
where $x=\omega/\omega_0$ is the frequency
normalized to the resonant frequency if the inclusion $\omega_0$
and $A,B$, and $C$ are constants.

In the optical literature, so called ``hierarchy of
polarizabilities" has been established for polarizabilities of
optical molecules \cite{raab}. According to that,
the strongest response is the electric dipole response (corresponding
to the electric polarizability $\alpha_{ee}$, the chirality parameter is
weaker, and the magnetic dipole response is the weakest of the
three considered here. This rule holds for usual molecules in the
optical region, when the molecules are small as compared with the
wavelength and spiral shapes are not very pronounced.

\begin{figure}[h]
\centering
\includegraphics[width=0.2\textwidth]{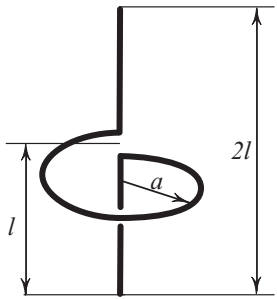}
\caption{The shape of the canonical helix, an illustration
for the antenna model of chiral particles \cite{antenna}.}
\label{helix}
\end{figure}

For our purpose we need a more general and quantitive model, that would describe
particles of arbitrary relations between the length of the
spiral and its diameter. To determine the relations between the
polarizabilities we make use of the analytical model \cite{antenna}
and calculate the ratios between the polarizabilities of small
canonical helices shown in Figure~\ref{helix}. The result is
\e \alpha_{em}=j\pi {a\over l}ka \, \alpha_{ee} ,\qquad \alpha_{mm}=
\left(\pi {a\over l}ka\right)^2\, \alpha_{ee} \l{good}\f
Here, $a$ is the loop radius and $l$ is the length of the dipole arm.
If $ka\ll 1$ and $\pi a/l\approx 1$, we come again to the simple
relations \r{hie}. However, for artificial chiral materials where
the proportions between the particle dimensions can be chosen at will,
and the loop radius is not always very small compared to the wavelength,
we should use the more general relations \r{good}.

Thus, we come to the following model for chiral particle polarizabilities:
\begin{equation}
 \alpha_{em}={jx\over{1-x^2+j\alpha x}},\qquad
\alpha_{ee}={\nu^{-1}\over{1-x^2+j\alpha x}},\qquad
\alpha_{mm}={\nu x^2\over{1-x^2+j\alpha x}}
\end{equation}
where we have denoted by $\nu=\pi (a/l) (ka)|_{\omega=\omega_0}$ the
coefficient which depends on the electrical diameter of the
spiral at resonance and on the ``form-factor'' (ratio of the
diameter and the length
of the helix). For natural optically active materials $\nu\ll 1$, but
for artificial chiral materials its value can be of the order of unity.

For the electric dipole inclusions we adopt the conventional
Lorentz dispersion formula, writing
\begin{equation}
\beta_{ee}={1\over{x_0^2-x^2+j\alpha' x}}
\end{equation}
Here  $x_0$ is the ratio between the resonant
frequency of the dipole particles and the resonant frequency of the helices.
In all expressions for the polarizabilities we omit constant (independent
from the frequency) amplitude coefficients, since they can be incorporated
into the number densities of chiral and dipole inclusions.

These models apply to inclusions whose dimensions are considerably
smaller than the wavelength. If for example the helix radius becomes
comparable to the wavelength, more complicated analytical and
numerical models should be
used (e.g. \cite{antenna,mariotte1}), but for the present purpose
we will not need more advanced models.

\section{Numerical examples}

\subsection{The role of resonant permittivity background}

\begin{figure}[h]
\centering
\includegraphics[width=0.50\linewidth]{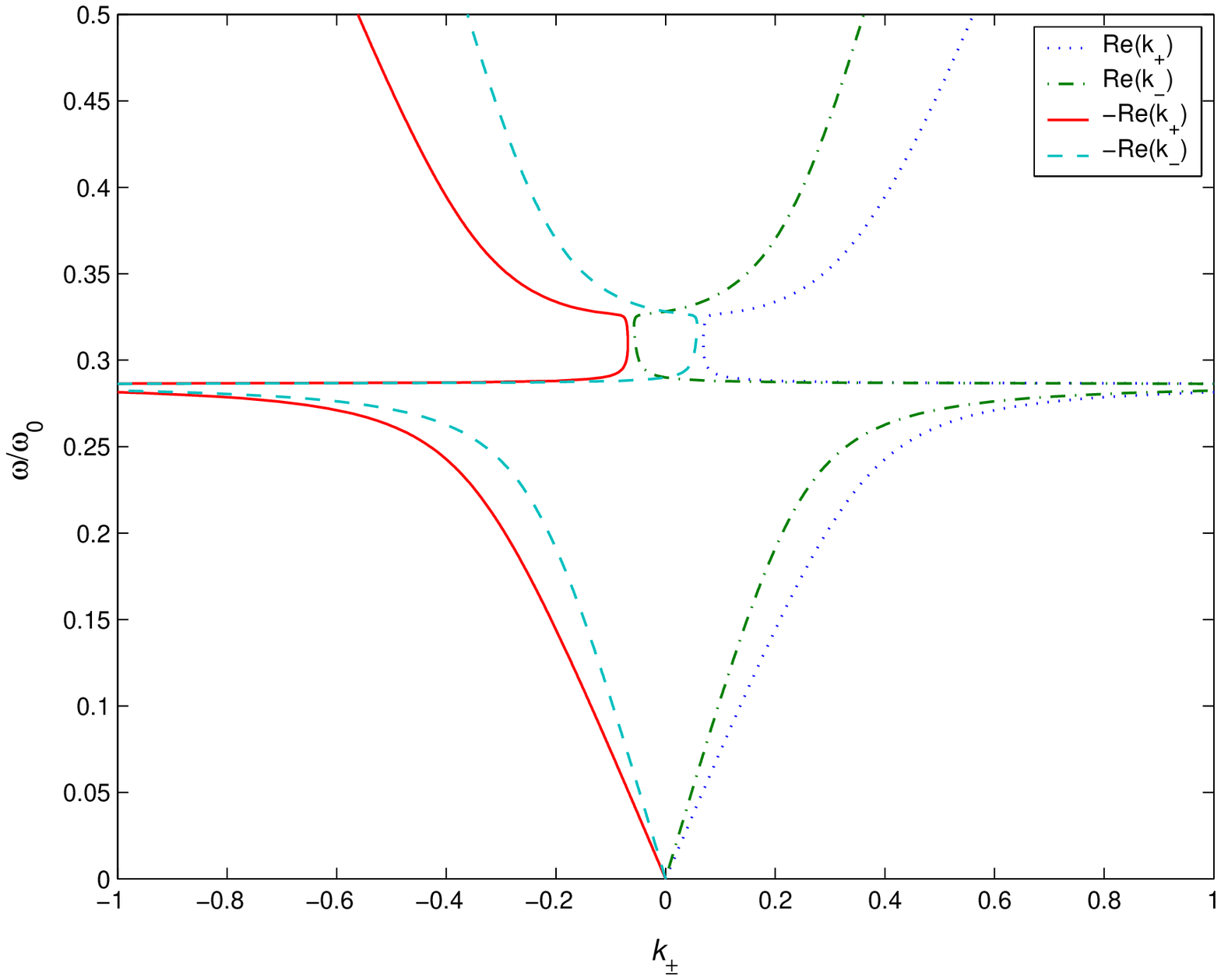}
\includegraphics[width=0.48\linewidth]{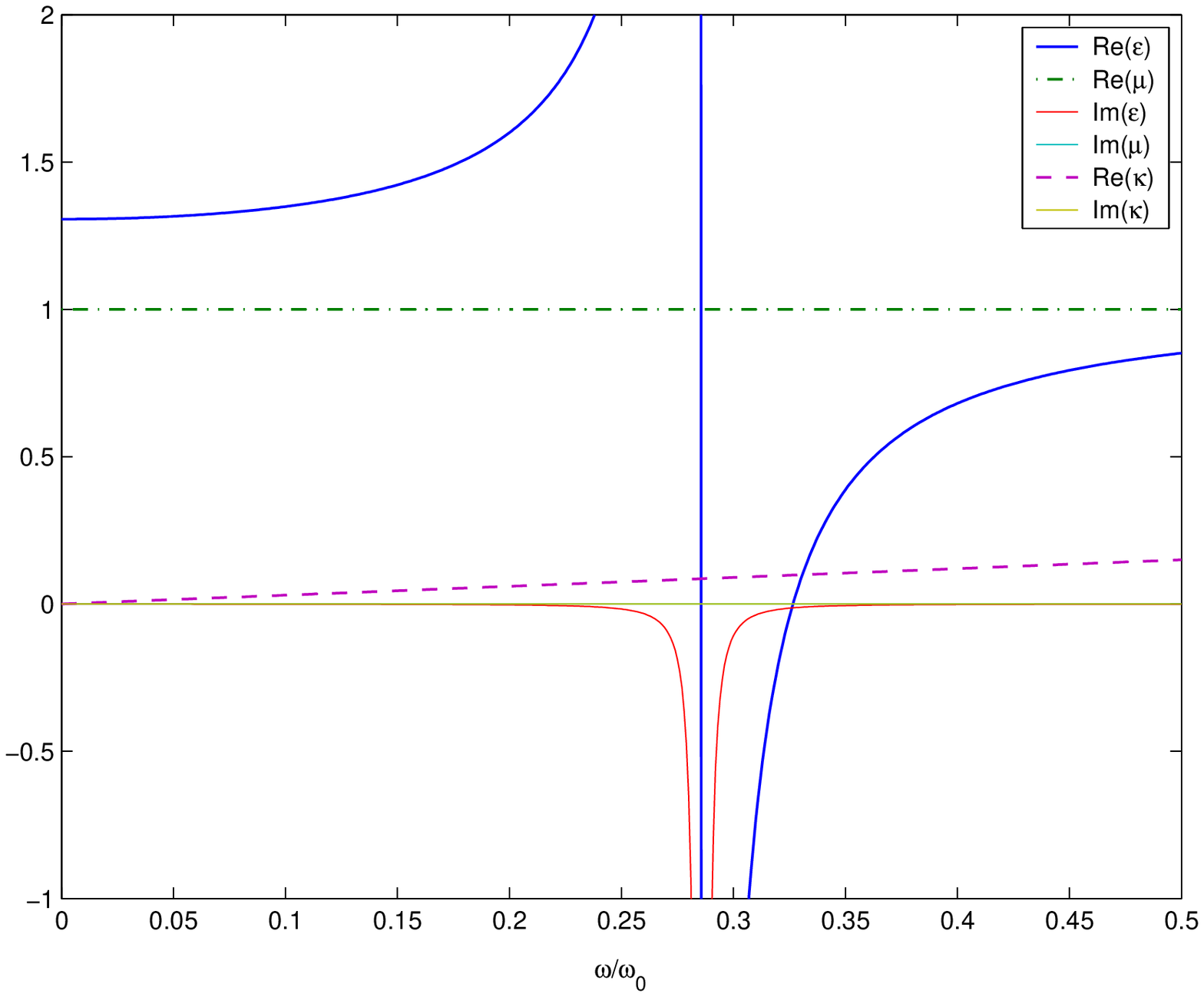}
 \caption{Dispersion curves and effective material parameters
 for the Pendry model of independent fractions.
 $\kappa=0.3\omega/\omega_0$. $n_2=0.025, \alpha_2=0.001$, resonant
 frequency of the dielectric phase is $0.3\omega_0$.}
\label{independent_fractions}
\end{figure}

Let us first calculate some numerical examples for using the simple model used in
paper \cite{pendry1}, where the chiral material was assumed to have no
frequency dispersion of the chirality parameter and the interaction between
two fractions was neglected. This approximation
corresponds to the frequency region well below the resonant frequency of
chiral inclusions (chirality parameter is still proportional to the frequency,
which we assume in our calculations,
but this dependence is not of principal importance for our problem).
Figure~\ref{independent_fractions} illustrates typical dispersion in this frequency range
for a material model of paper \cite{pendry1}.
In this example, we have taken a very high chirality parameter,
to highlight the effect of chirality (in real composites, the effect is usually
pretty weak at frequencies well below the resonance). The chiral fraction
has been assumed to be lossless.
We observe that near the resonant frequency of dipole inclusions there
is a very narrow backward-wave band. This effect happens when the permittivity
is very small but still positive. Closer to the resonant frequency of dipoles there is a
wide stop band.

However, the model used in this calculation is not realistic, because
near the resonant frequencies all the material parameters resonate due to
strong field interactions between particles of two phases (see Section~\ref{mixing}).
Next, we calculate an example using the model of the mixture and
the particles polarizabilities introduced above.


\begin{figure}[h]
\centering
\includegraphics[width=0.50\linewidth]{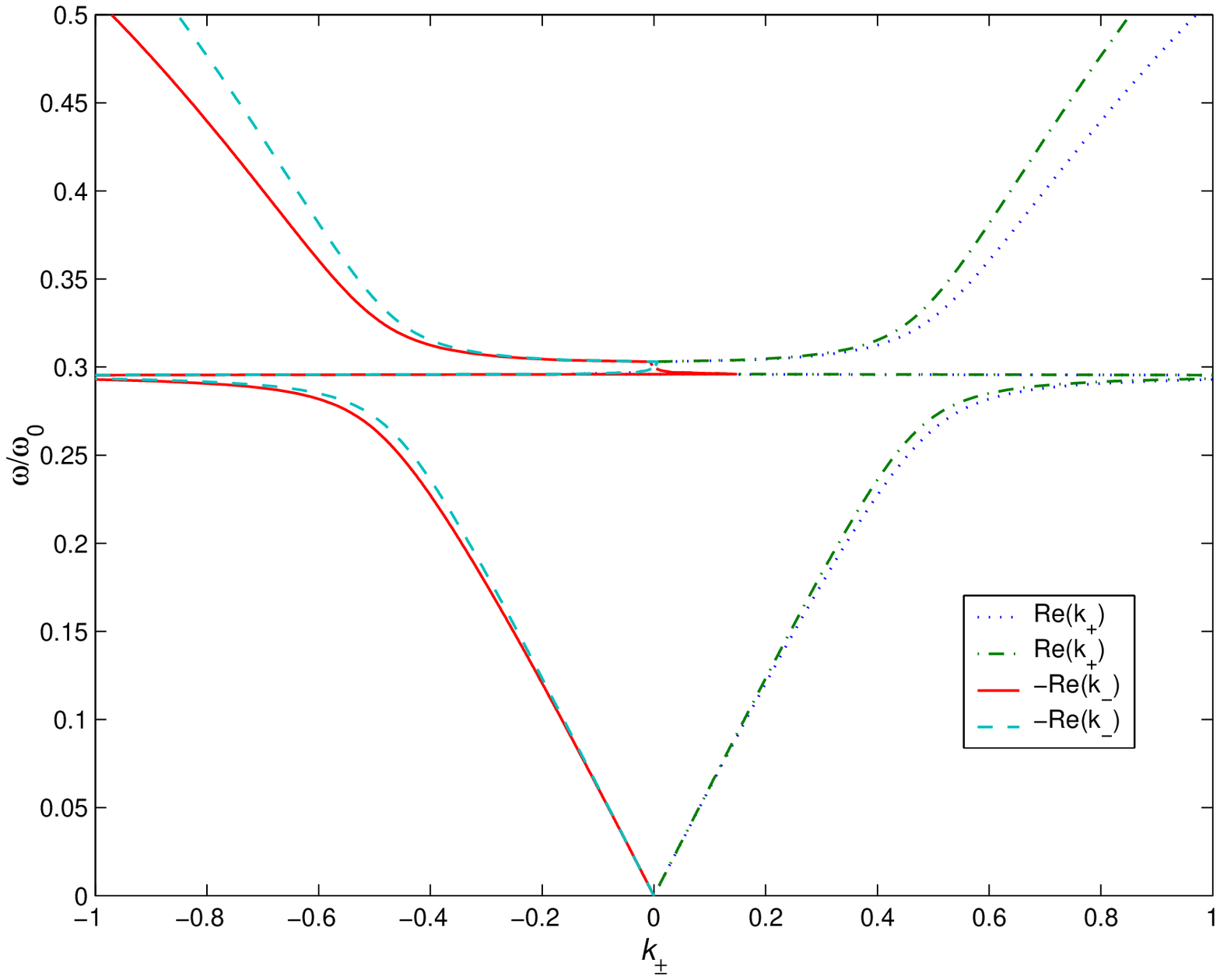}
\includegraphics[width=0.48\linewidth]{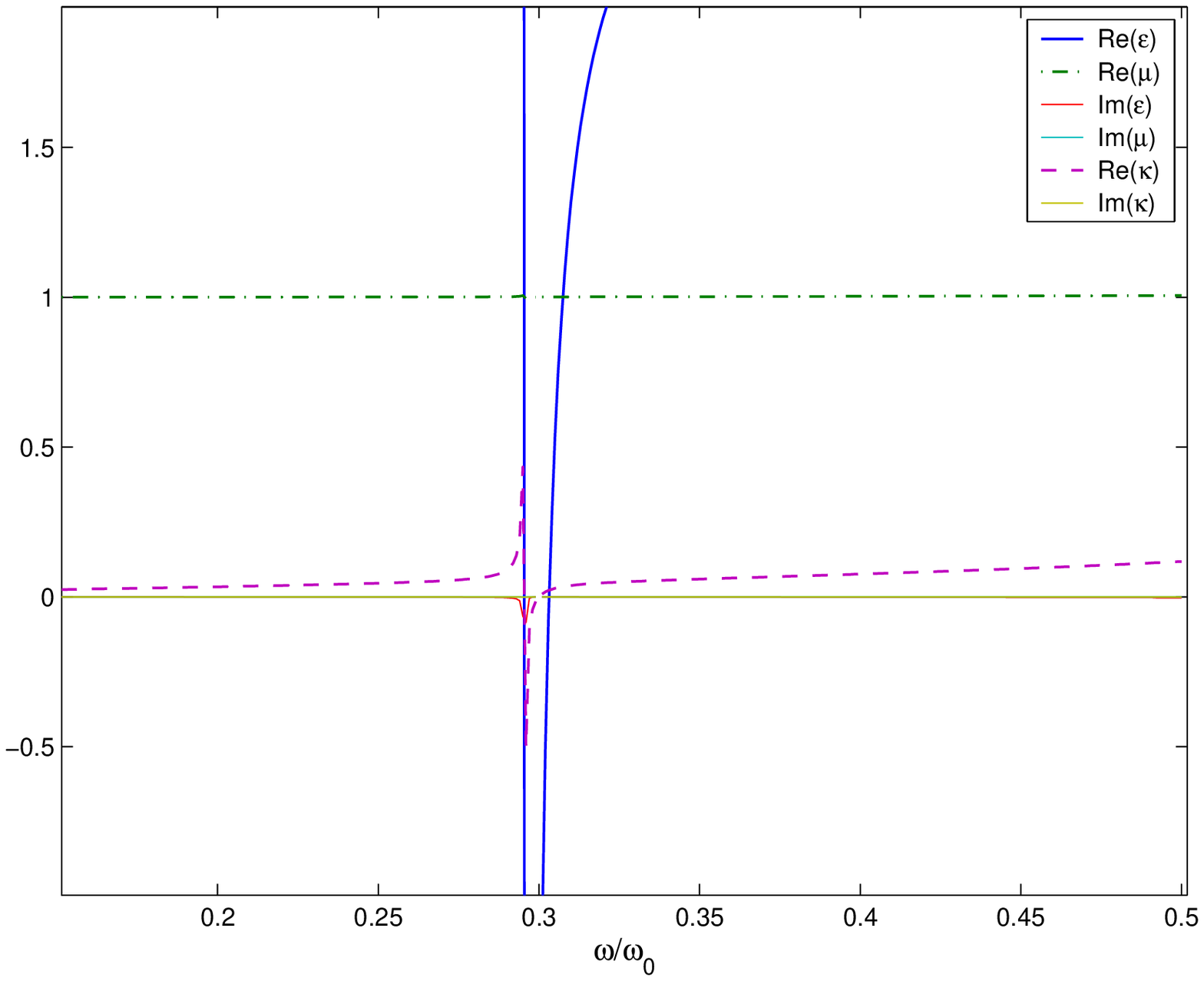}
\caption{Dispersion curves for a chiral medium with resonant electric dipole inclusions
and its effective material parameters.
The frequency is normalized to the
resonant frequency $\omega_0$.}
\label{mixture}
\end{figure}


\begin{figure}[h]
\centering
\includegraphics[width=0.6\linewidth]{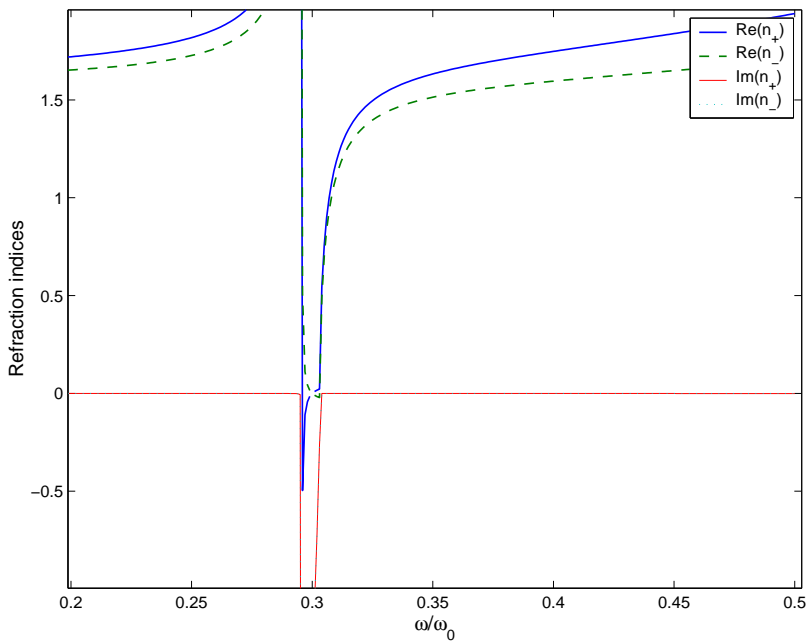}
\caption{Negative and imaginary parts of the refraction indices of
two eigenmodes in the same medium as in Figure~\ref{mixture}.}
\label{indices}
\end{figure}

The resonant frequency of the electric dipole fraction is considerably lower than the resonant
frequency of the helices of the chiral mixture ($0.3\omega_0$).
The following parameters of the medium have been assumed:
$n_1=0.1$, loss factor $\alpha=0.001$, $n_2=0.05$, loss factor of the electric fraction
particles $\alpha'=0$. Parameter $\nu=0.1$. In order to
achieve a backward-wave effect, we assumed a very high
concentration of chiral particles (in naturally available
optical chiral materials the chirality parameter is usually orders of
magnitude smaller) and assume that the electric dipole particles are
lossless. The results are shown in Figure~\ref{mixture}.
 We see that even under the above assumptions, there is actually no backward-wave band.
This is apparent from the plot of the refractive indices of the
two eigenmodes. The narrow frequency band where the real part of one of the indices
becomes negative is already inside the stop band.

Near the resonance of the electric phase all the material parameters
also resonate. The resonant increase of the chirality parameter looks like a
possibility to satisfy the backward-wave condition \r{condition},
because in the frequency band where the chirality parameter increases, the permittivity
takes small
values. However, this does not lead to a backward-wave regime.
The problem is that in these two-phase composites
with resonant electric dipoles embedded in a natural optically
active material the effective magnetic permeability
resonates very weakly. In the resonant band of the permittivity
the effective permeability still stays near unity,
and there is a wide stop band near the resonance of the
electric phase. We have also studied the situation when the
magnetic response of helical fraction is stronger, taking the value of
parameter $\nu$ to be equal unity. This does not change the
above conclusion, because the magnetic response considerably increases
only near the resonance of the helix.

\subsection{Interplay of the resonances of helices and dipoles}

To achieve strong chiral response in practice, the working
frequency should be close to the resonant frequency of chiral inclusions.
Interplay of two phases becomes quite strong in this situation, and the
mixture shows rather complicated frequency response. This is illustrated by
numerical examples in this section.

\begin{figure}[h]
\centering
\includegraphics[width=0.48\linewidth]{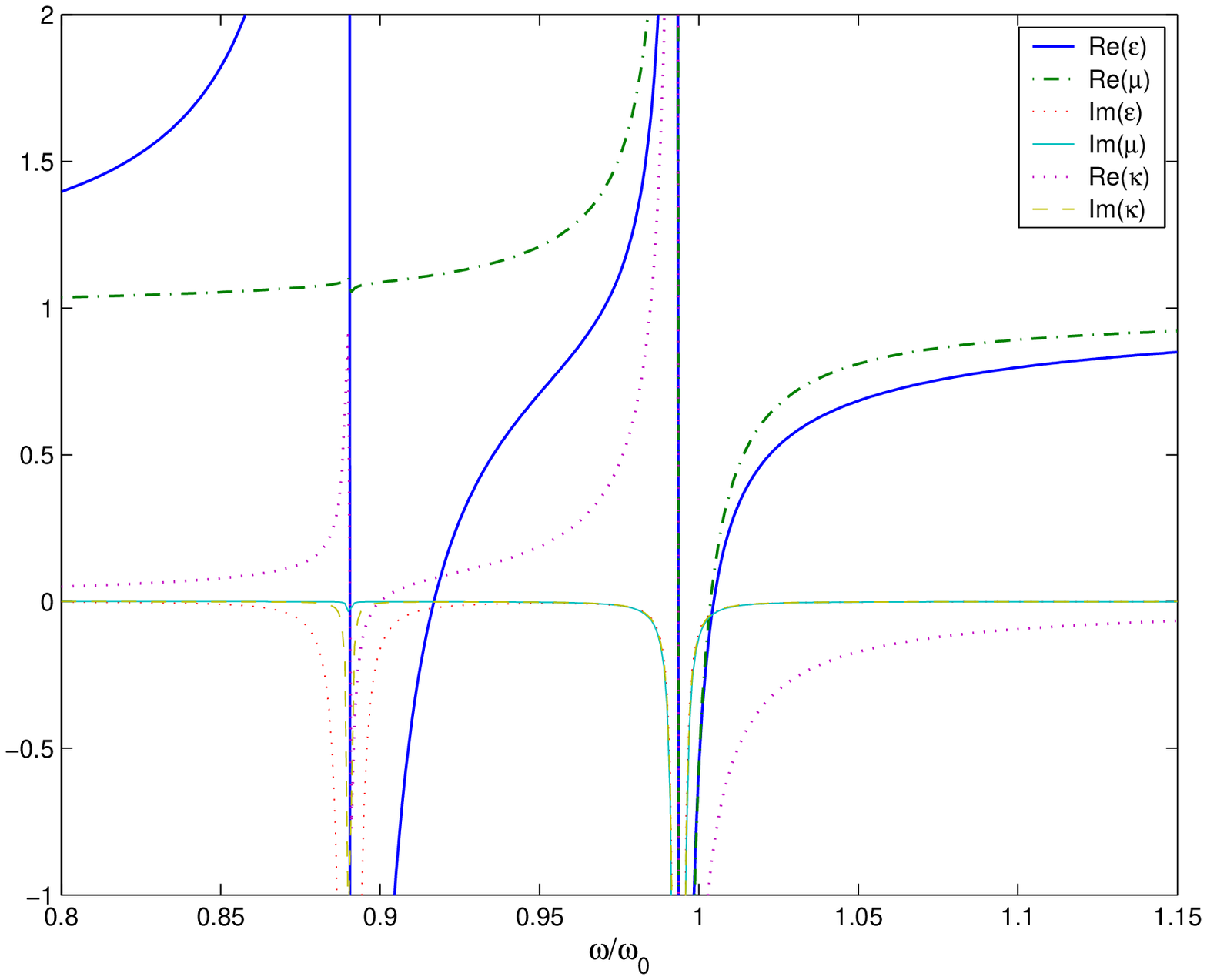}
\includegraphics[width=0.50\linewidth]{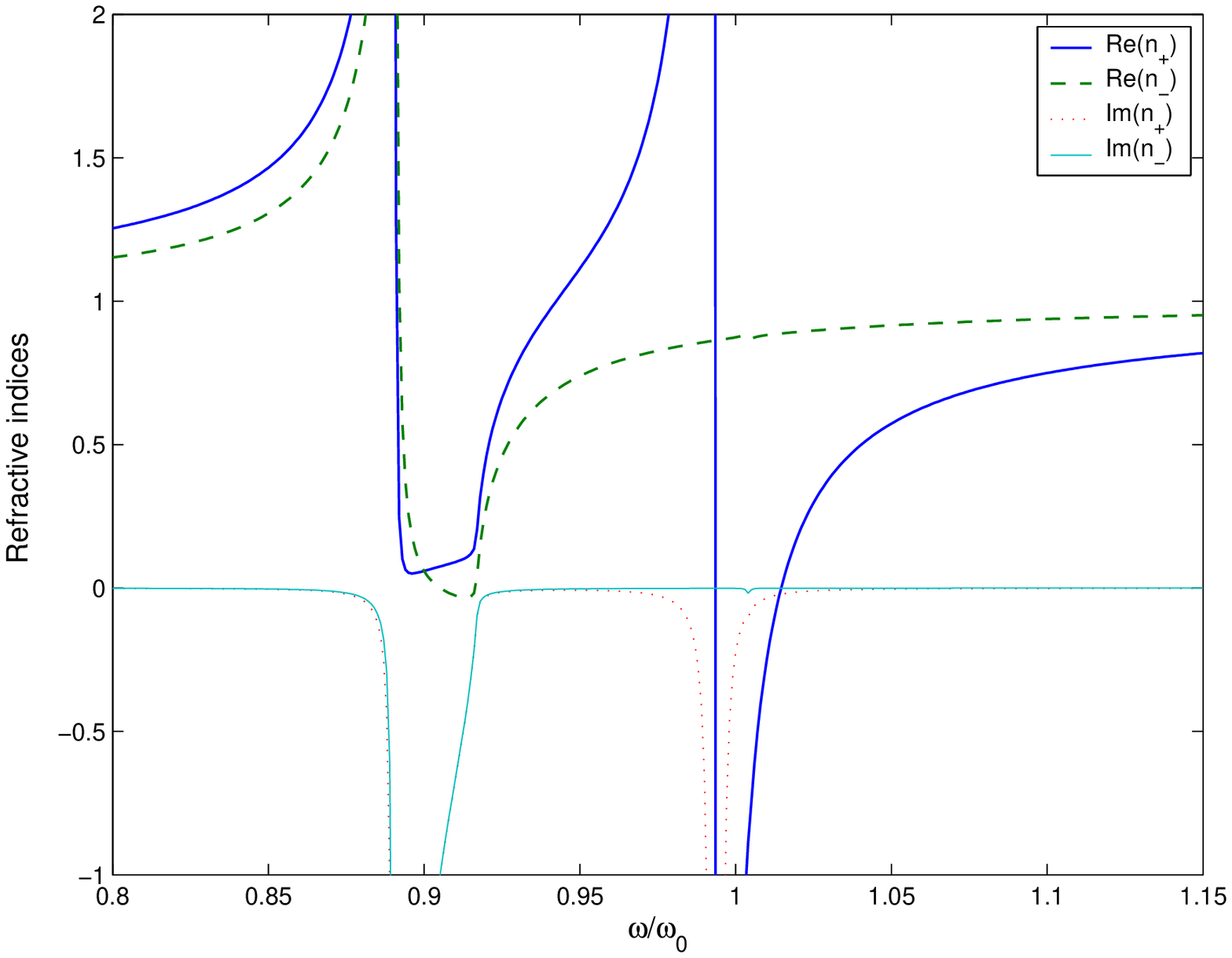}
 \caption{Effective material parameters and refraction indices for a mixture of dipoles and
 helixes with the parameters $n_1=0.02$, $n_2=0.05$,
 $\alpha=\alpha'=0.001$, and $\nu =1$. The resonant frequency of
 electric dipoles equals $0.9\omega_0$.}
\label{fig_param}
\end{figure}

An example of frequency dependence of effective material parameters and refraction
indices is shown in
Figure~\ref{fig_param}. The two phases interact strongly and all the
material parameters show resonant response near the resonant
frequencies of helices and dipoles. Although we assumed that helices
have strong magnetic properties ($\nu=1$), we see that predicted in
\cite{pendry1} backward-wave frequency band
near the resonance of dipoles falls into a stop band of the mixture.
Close
to the resonance frequency of the effective permittivity,
$\mu_{\rm eff}$ is nearly unity and $\kappa_{\rm eff}$ also has a
resonance because of coupling between dipoles and helices.
It is difficult to design a backward-wave material
to operate near the resonance of background permittivity
because the resonances of the effective permittivity and
permeability can not be tailored separately.

On the other hand, we observe that there indeed exists predicted in
\cite{bokut,sergei1} a backward-propagation band
near the resonant frequency of helices.

%

\subsection{Negative refraction in resonant chiral composites}

\begin{figure}[h]
\includegraphics[width=0.49\linewidth]{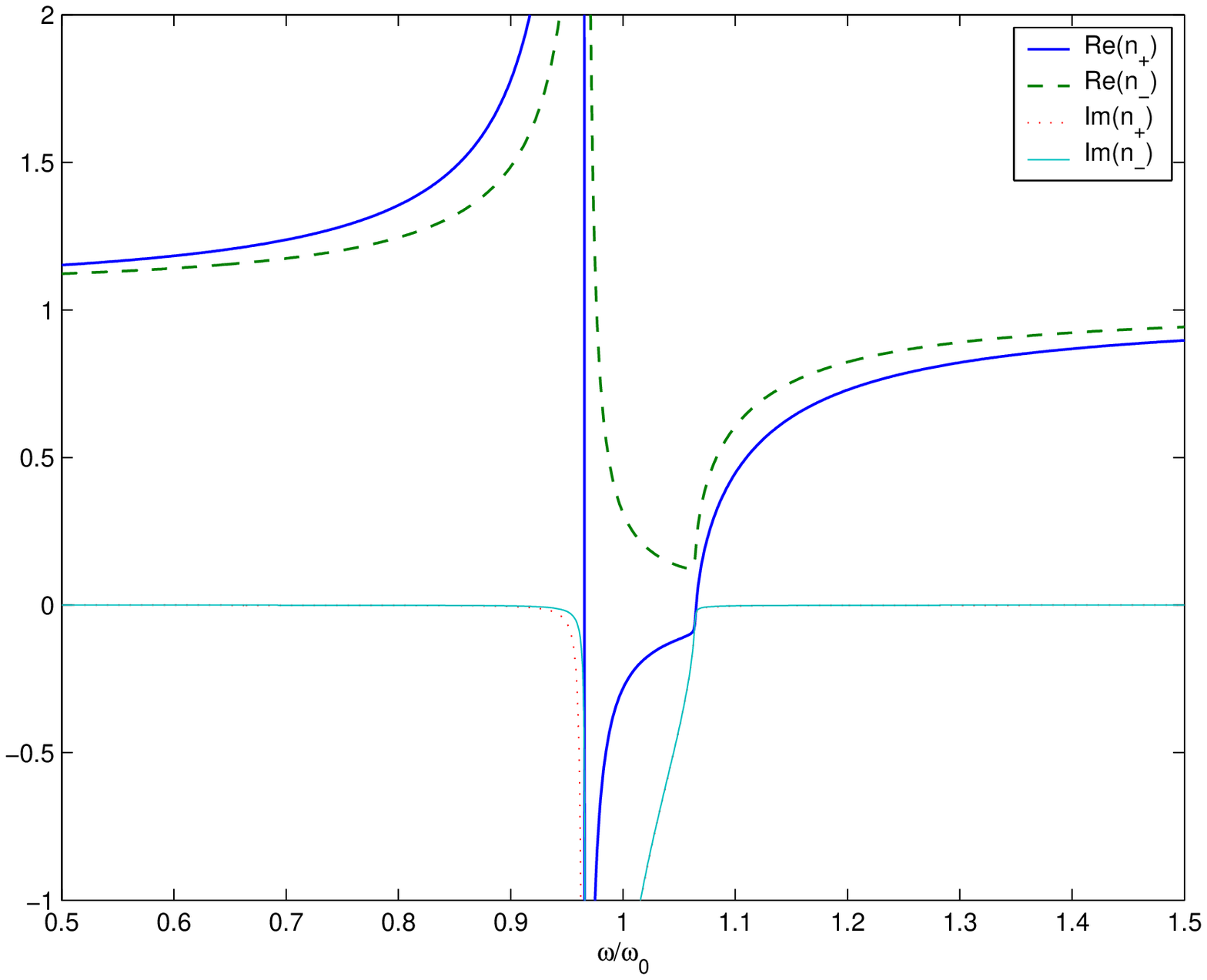}
\includegraphics[width=0.49\linewidth]{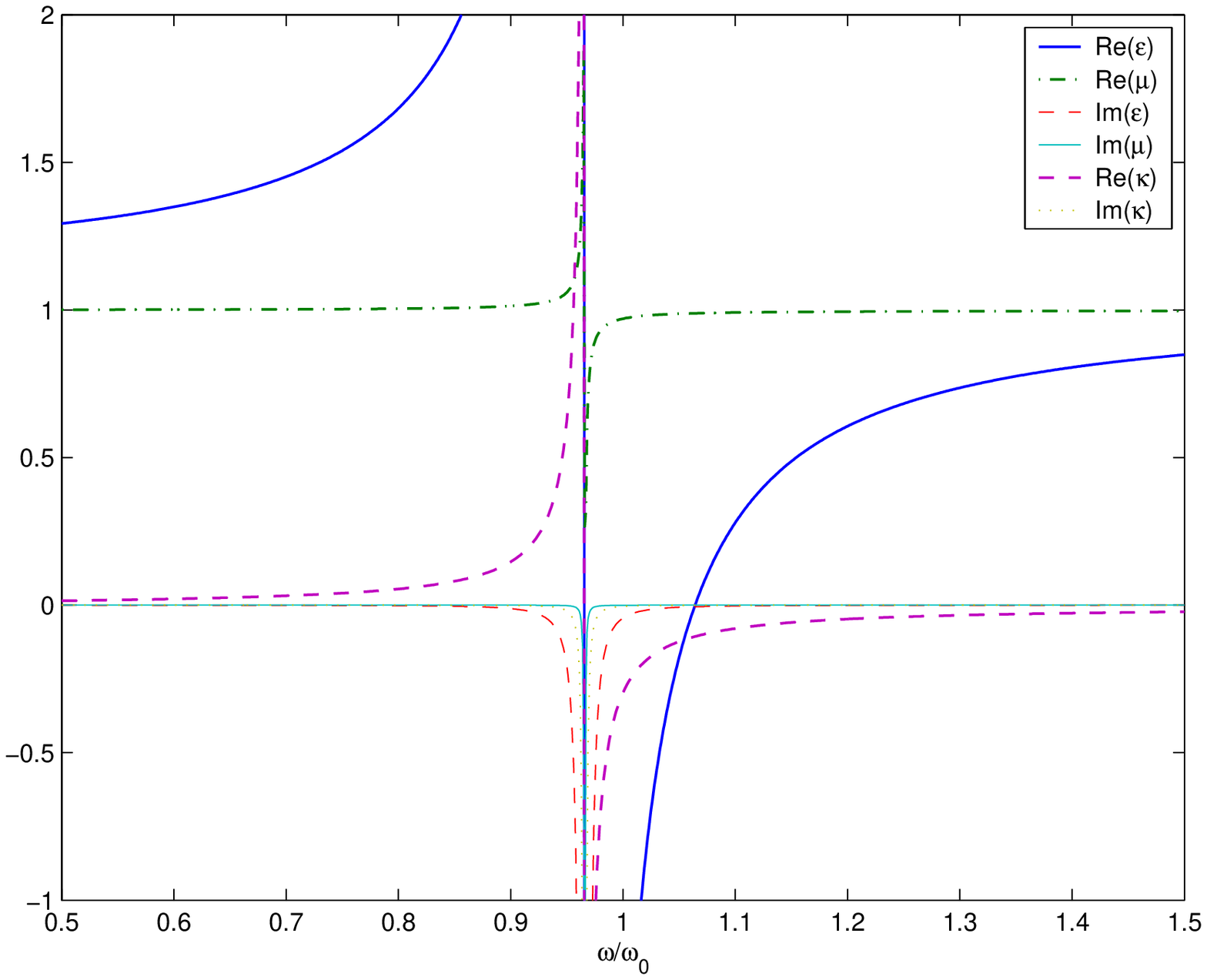}
\caption{Refractive indices and effective parameters of
a chiral material with the parameters $n_1=0.02$, $\alpha=0.001$, $\nu =0.1$.}
\label{chiral_resonant}
\end{figure}

The results of the previous section indicate that the most
appropriate approach to realize backward-wave materials and superlenses
with the use of chiral materials is the use of only chiral inclusions.
In this section we give two examples of such mixtures to
illustrate the design criteria for these materials.
Figure~\ref{chiral_resonant} gives an example of a resonant chiral
material with the parameters that are probably possible to
achieve using natural materials (parameter $\nu$ is much smaller than unity
and magnetic properties are weak, even close to the particle resonance).
This is the case first considered in \cite{bokut}. The results
shown in
Figure~\ref{chiral_resonant} lead to the conclusion that also here
there is no propagating backward-wave regime --- when the
real part of the refraction
index becomes negative, its imaginary part is very large.

\begin{figure}[h]
\includegraphics[width=0.49\linewidth]{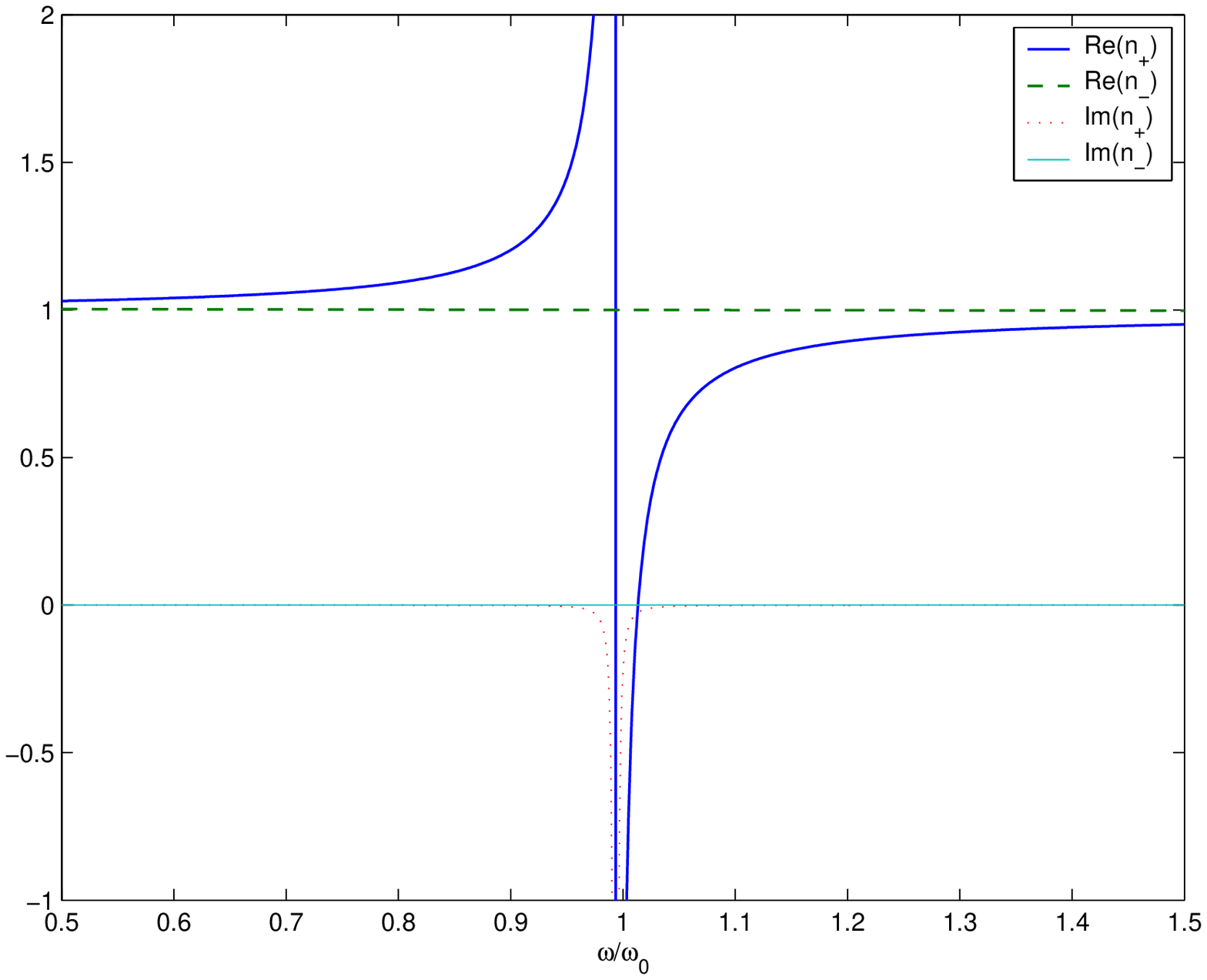}
\includegraphics[width=0.49\linewidth]{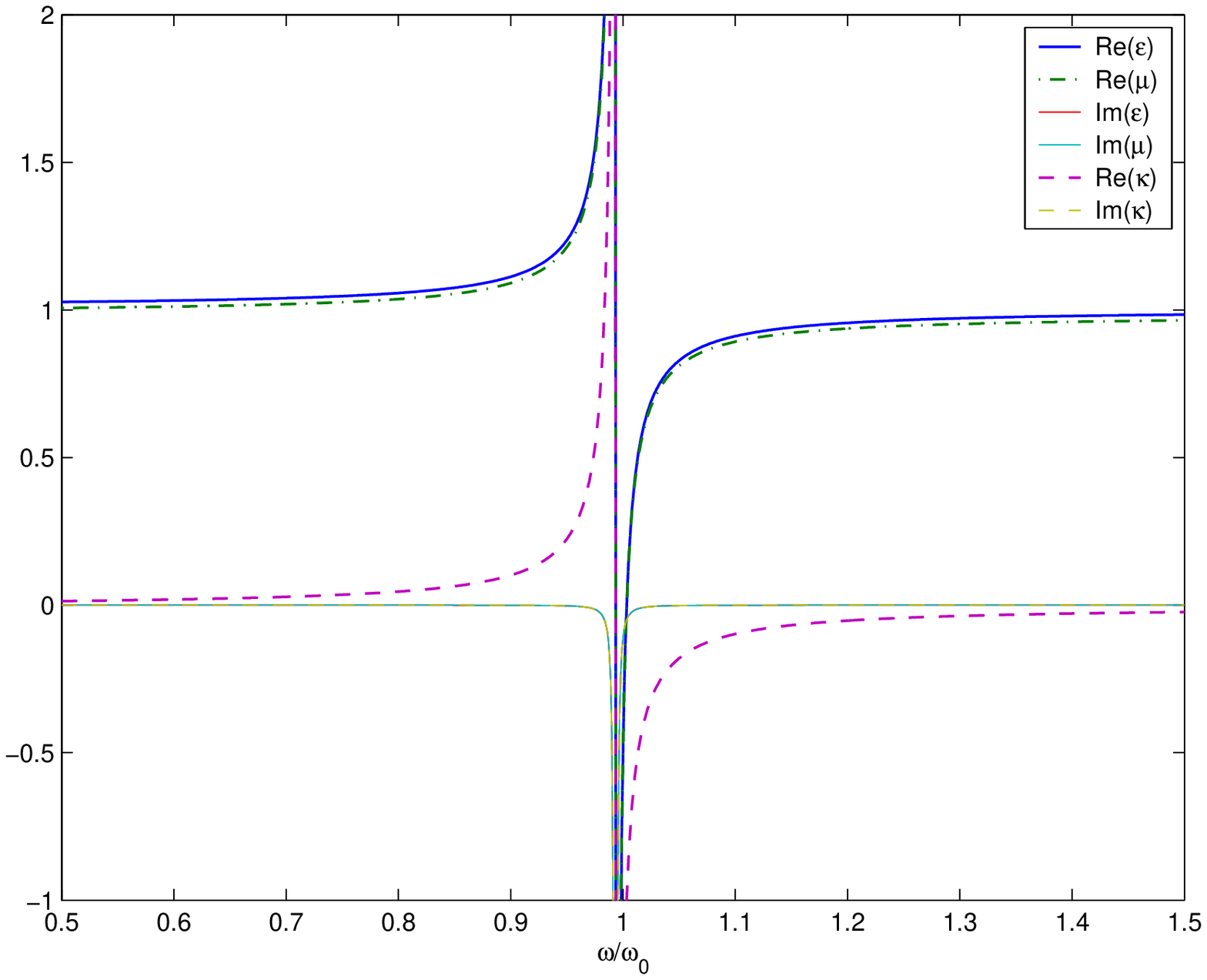}
\caption{Refractive indices and effective parameters of
a ``chiral nihility" composite material with the parameters $n_1=0.02$, $\alpha=0.001$, $\nu=1$.}
\label{nihility}
\end{figure}

The last Figure \ref{nihility} illustrates the situation considered first in
\cite{sergei1}. In this case we chose a large value of parameter $\nu=1$.
The effective permeability is now much stronger due to a larger ratio of the
helix diameter to its length $a/l$ and (or) the electric diameter of the
helix $ka$. As a result, the permittivity and permeability
follow nearly the same dispersion laws, and the real part of their product
remains positive even when the permittivity goes into negative.
In the previous example there is a stop band in this frequency area,
but in the present situation there is a propagating backward-wave regime like
in the usual double-negative materials based on for example split rings and
wire lattices. The role of chirality parameter here is two-fold. First,
is the chirality parameter is non-zero, this backward-wave band is wider,
since its limits are given by a weaker condition \r{condition} than the
usual ${\rm Re}\sqrt{\epsilon\mu}<0$. Second, the possible physical effects in
this new material are reacher than in double-negative materials (see \cite{sergei1}).
However, such materials do not exist in nature and have to be manufactured
artificially as {\em metamaterials}. This is reasonably easy for
microwave frequencies, and there are many reports in the literature
about artificial chiral media with strong resonant response (although
we are not aware about any realization of the backward-wave regime).
For the optical frequency range, the small required particle size
makes the task very challenging, but we hope that the quickly developing
nano-technologies will make it reality.

\section{Conclusions}

Possibilities to achieve negative refraction and enhancement of
evanescent fields in a ``perfect lens'' using chiral materials
have been theoretically explored. The two known variants have been considered:
the use of a mixture of
helixes and resonant dipoles \cite{pendry1} and the use of a composite of
only helices \cite{bokut,sergei1}.
Mixing equations for the effective material parameters, which take
into account the coupling between dipoles and helixes has been given.
Numerical examples have been calculated with the use of an introduced
general dispersion law for the polarizabilities of helical particles.
This model gives a possibility to study particles with different
shapes and electrical sizes.

It has been shown that once the coupling between helices and dipoles
is taken into account, there
is a stop band in the frequency region where negative refraction
was expected to occur. However, the negative refraction can still occur in
a ``chiral nihility'' materials as has been suggested in \cite{sergei1}.
Negative refraction could exist at frequencies higher than the resonant frequency of
chiral particles. The role of chirality is seen in widening the
backward-wave frequency band and in opening a way to realize new
physical effects and possibly create new microwave and optical devices.
Realization of these new media for optical applications requires
manufacturing of chiral inclusions with controllable shape
that would exhibit resonant response in the optical regime.

\end{document}